\documentclass[12pt]{article}

\usepackage{graphicx}
\begin{document}

\begin{center}
{\bf Black hole as a magnetic monopole within exponential nonlinear electrodynamics } \\
\vspace{5mm} S. I. Kruglov
\footnote{E-mail: serguei.krouglov@utoronto.ca}
\underline{}
\vspace{3mm}

\textit{Department of Chemical and Physical Sciences, University of Toronto,\\
3359 Mississauga Road North, Mississauga, Ontario L5L 1C6, Canada} \\
\vspace{5mm}
\end{center}

\begin{abstract}
We perform the gauge covariant quantization of the exponential model of nonlinear electrodynamics.
Magnetically charged black holes, in the framework of our model are considered, and the regular black hole
solution is obtained in general relativity. The asymptotic black hole solution at $r\rightarrow \infty$ is found.
We calculate the magnetic mass of the black hole and the metric function which are expressed via the parameter $\beta$ of the model and the magnetic charge. The thermodynamic properties
and thermal stability of regular black holes are analysed. We calculate the Hawking temperature of black holes and their heat capacity at the constant magnetic charge. We find a point where the temperature changes the sign that corresponds to the first-order phase transition. It is shown that at critical point, where the heat capacity diverges, there is a phase transition of the second-order.
We obtain the parameters of the model when the black hole is stable.
\end{abstract}

\section{Introduction}

Classical electrodynamics is modified due to quantum corrections and becomes nonlinear electrodynamics (NLED). Thus, one-loop corrections
in QED lead to nonlinear Heisenberg-Euler electrodynamics \cite{Heisenberg} which admits the phenomenon of vacuum birefringence. This effect,
when indexes of refraction in the presence of the external magnetic field depend on polarization states, is now under experimental verification by PVLAS and BMV collaborations. Therefore, viable models of NLED should describe the birefringence phenomenon.
In well-known Born-Infeld electrodynamics \cite{Born} the effect of birefringence is absent.
But in the modified Born-Infeld electrodynamics, containing two parameters, the birefringence phenomenon takes place \cite{Krug}.
Another requirement for NLED, to be a viable model, is that for weak fields NLED has to approach to Maxwell's electrodynamics.
For strong electromagnetic fields classical electrodynamics can be modified because of the self-interaction of photons \cite{Jackson}.
In Born-Infeld electrodynamics and in some models of NLED \cite{Shabad} - \cite{Kruglov7} there is an upper bound on the electric field in the
centre of charged particles and the total electromagnetic energy is finite. In Maxwell's electrodynamics
the problem of singularity of an electric field at the origin of charged  particles and the problem of infinite electromagnetic energy exist. These problems can be solved in NLED.
In addition, NLED coupled with general relativity (GR) can result in the universe acceleration \cite{Garcia} - \cite{Kruglov4}. At the same time, electromagnetic fields in Born-Infeld electrodynamics do not undergo to the universe acceleration \cite{Novello1} and Born-Infeld electrodynamics
 possesses the problem of causality \cite{Quiros}.
The black hole solutions in GR within NLED were studied in \cite{Breton} - \cite{Kruglov8}, and it was shown that
these solutions approach to the Reissner-Nordstr\"{o}m (RN) solution at $r\rightarrow \infty$.

In this paper we consider exponential nonlinear electrodynamics, proposed in \cite{Kruglov0}, coupled to GR.
We investigate magnetically charged black holes and obtain solutions similar to RN solution with some corrections. The thermodynamics of such
black holes is also studied. We demonstrate that there are the first-order and the second order phase transitions in black holes.

The structure of the paper is as follows. It was shown in Section 2 that causality and unitarity principles are satisfied
in our model. We perform the gauge covariant Dirac quantization of the exponential electrodynamics.
NLED coupled with GR is investigated in Section 3 and we obtain black hole solutions. We find corrections to RN solution at $r\rightarrow\infty$. In this Section the magnetic mass and metric function are obtained. We show that weak energy condition holds in exponential nonlinear electrodynamics.
It is demonstrated that only at $b=2^{3/2}\sqrt{\beta}/(qG)\leq0.83$ there are two or one horizons.
We show that there are not singularities of Ricci's scalar at $r\rightarrow \infty$ and at $r\rightarrow 0$ .
Thermodynamics of black holes is studied in Section 4. It is shown that in black holes the first-order and second-order phase transitions take place.
In Section 5 we discuss the results obtained. In Appendix we estimate the Kretschmann scalar and its asymptotic.

The units $c=\hbar=1$ and the metric tensor signature $\eta=\mbox{diag}(-1,1,1,1)$ are used. The Greek indexes take values $0,1,2,3$, while for the spatial indexes, designated by the Latin letters, the values are $1,2,3$.

\section{The gauge covariant Dirac quantization of exponential electrodynamics}

Let us consider exponential nonlinear electrodynamics proposed in \cite{Kruglov0}.
The Lagrangian density of exponential electrodynamics reads
\begin{equation}
{\cal L} = -{\cal F}\exp(-\beta{\cal F}),
 \label{1}
\end{equation}
where ${\cal F}=(1/4)F_{\mu\nu}F^{\mu\nu}=(\textbf{B}^2-\textbf{E}^2)/2$, $F_{\mu\nu}=\partial_\mu A_\nu-\partial_\nu A_\mu$.
The parameter $\beta$ possesses the dimension of the (length)$^4$ and the upper bound on the $\beta$ ($\beta \leq 1\times 10^{-23}$ T$^{-2}$) was found from PVLAS experiment \cite{Kruglov0}. The field equations follow from the Lagrangian density (1) \cite{Kruglov0}
\begin{equation}
\nabla\cdot \textbf{D}= 0,~~~~ \frac{\partial\textbf{D}}{\partial
t}-\nabla\times\textbf{H}=0,
\label{2}
\end{equation}
where the electric displacement field is given by
\begin{equation}
\textbf{D}=\textbf{E}\left(1-\beta {\cal F}\right)\exp(-\beta{\cal F}),
\label{3}
 \end{equation}
and the magnetic field is
\begin{equation}
\textbf{H}= \textbf{B}\left(1-\beta {\cal F}\right)\exp(-\beta{\cal F}).
\label{4}
\end{equation}
Equations of vacuum nonlinear electrodynamics correspond to the continuous media electrodynamics equations with the specific
constitutive relations (3) and (4). The second pair of the Maxwell equations, that is the consequence of the Bianchi identity, reads
\begin{equation}
\nabla\cdot \textbf{B}= 0,~~~~ \frac{\partial\textbf{B}}{\partial
t}+\nabla\times\textbf{E}=0.
\label{5}
\end{equation}
The theory is viable if general principles of causality and unitarity are satisfied. The causality principle tells us that the group velocity of excitations over the background should be less than the light speed. This gives the requirement
$ {\cal L}_{\cal F}\equiv \partial{\cal L}/\partial{\cal F}\leq 0$ \cite{Shabad2}. We find from Eq. (1)
\begin{equation}
{\cal L}_{\cal F}=-( 1-\beta{\cal F})\exp(-\beta{\cal F}).
\label{6}
\end{equation}
Thus, at $\beta{\cal F}\leq 1$  the causality principle holds and tachyons will not appear. For pure magnetic field ($\textbf{E}=0$), which will be considered, this requirement is $B\leq \sqrt{2}/\sqrt{\beta}$.
The unitarity principle requires
${\cal L}_{\cal F}+2{\cal F}{\cal L}_{{\cal F}{\cal F}}\leq 0$ and ${\cal L}_{{\cal F}{\cal F}}\geq 0$ \cite{Shabad2}. From Eq. (1) one obtains
\[
{\cal L}_{{\cal F}{\cal F}}= \beta(2- \beta{\cal F})\exp(-\beta{\cal F}),
\]
\begin{equation}
{\cal L}_{\cal F}+2{\cal F}{\cal L}_{{\cal F}{\cal F}}=(-1+5\beta{\cal F}-2\beta^2{\cal F}^2)\exp(-\beta{\cal F}).
\label{7}
\end{equation}
We find from Eqs. (7) that the unitarity principle gives the restriction $\beta{\cal F}\leq (5-\sqrt{17})/4\simeq0.219$.  
The violation of fundamental principles might signal some inconsistencies in the theory.
The unitarity principle guaranties that the residue of the photon propagator is nonnegative. In other words the
norm of every elementary excitations has to be positive. In this case there will be no ghosts.
As a result, causality and unitarity of the theory take place at $\beta{\cal F}\leq (5-\sqrt{17})/4$. For the case $\textbf{E}=0$ this gives $ B\leq \sqrt{(5-\sqrt{17})/(2\beta)}\simeq 0.66/\sqrt{\beta}$.

The Lagrangian density (1) is invariant under gauge transformations described by the $U(1)$ group. The phase space, as in any field theory, is infinite dimensional. The Lagrangian, corresponding to Lagrangian density (1), is given by $L=\int d^3x {\cal L}$, and the action is $I=\int dt L$. We will study the time evolution of fields, and therefore, the formalism looks like non-relativistic although the theory is Lorentz covariant.
The ``coordinates" and ``velocities" here are $A_\mu$ and $\partial A_\mu/\partial t\equiv \partial_0 A_\mu$, respectively.
From Eq. (1) according to the Dirac procedure  \cite{Dirac} we find the momenta
 \begin{equation}
 \pi_i=\frac{\partial {\cal L}}{\partial(\partial_0 A^i)}=-E_i( 1-\beta{\cal F})\exp(-\beta{\cal F})=-D_i,~~~
 \pi_0=\frac{\partial {\cal L}}{\partial (\partial_0 A^0)}=0.
\label{8}
\end{equation}
Thus, the spatial part of the momentum $\pi$ equals the displacement field $\textbf{D}$ with the opposite sign.
From second equation in (8) one finds the primary constraint
\begin{equation}
 \varphi_1 (x)\equiv\pi_0 ,\hspace{0.3in}\varphi_1 (x)\approx 0.
\label{9}
\end{equation}
We use Dirac's symbol $\approx$ for the equation that holds
only weakly, i.e. $ \varphi_1 (x)$ can possess nonzero Poisson brackets with some variables.
Equations (9) represent an infinite set of constraints for every spatial coordinate $\textbf{x}$.
With the help of the Poisson brackets $\{A_i,\pi_j\}=\delta_{ij}\delta(\textbf{x}-\textbf{y})$, and using equation $\pi_j=-D_j$, we arrive at
\begin{equation}
 \{A_i (\textbf{x},t),D_j(\textbf{y},t)\}=-\delta_{ij}
 \delta(\textbf{x}-\textbf{y}).
\label{10}
\end{equation}
Multiplying Eq. (10) by the the operator $\epsilon^{mki}\partial/\partial x^k$ ($\epsilon^{mki}$ is the antisymmetric Levi-Civita symbol) and performing a summation over repeated indexes, one obtains the
Poisson brackets between the magnetic induction field $\textbf{B}=\nabla\times \textbf{A}$ and the electric displacement field $\textbf{D}$
\begin{equation}
 \{B^m (\textbf{x},t),D^j(\textbf{y},t)\}=\epsilon^{mjk}\frac{\partial}{\partial x^k }
 \delta(\textbf{x}-\textbf{y}).
\label{11}
\end{equation}
Equation (11) also takes place in Born-Infeld electrodynamics \cite{Born1}, \cite{Dirac}.
In the quantum field theory, the Poisson brackets should be replaced by the quantum commutator,
$\{B,D\}\rightarrow -i\left[B,D\right]$, where $\left[B,D\right]=BD-DB$. With the help of Eqs. (1) and (8),
and the relation ${\cal H}=\pi^\mu\partial_0 A_\mu-{\cal L}$, we find the Hamiltonian density:
\begin{equation}
 {\cal H}=D_iE^i+{\cal F}\exp(-\beta{\cal F})+\pi^m \partial_m A_0.
\label{12}
\end{equation}
Because the primary constraint (9) should be a constant of motion, we arrive at the equation
\begin{equation}
 \partial_0 \pi_0 =\{\pi_0,H\}=\partial_m \pi^m= 0.
\label{13}
\end{equation}
Here $H=\int d^3x {\cal H}$ is the Hamiltonian. Equation (13) guarantees that the primary constraint (9) is conserved and represents the Gauss law as $\pi_i=-D_i$.
From Eq. (13) we arrive at the secondary constraint
\begin{equation}
 \varphi_2 (x)\equiv\partial_m \pi^m ,\hspace{0.3in}\varphi_2 (x)\approx 0.
\label{14}
\end{equation}
Both constraints (9) and (14) can be considered on the same footing.
We note that the weak equality $\approx$ is not compatible with the Poisson brackets \cite{Dirac}.
The time evolution of the second constraint is given by
\begin{equation}
\partial_0 \varphi_2 =\{\varphi_2,H\}\equiv 0.
\label{15}
\end{equation}
Equation (15) shows that there are no additional constraints. Because $\{\varphi_1,\varphi_2\}=0$ there are not second class constraints.
In Maxwell's electrodynamics and NLED \cite{Dirac}, \cite{Kruglov9}, \cite{Kruglov7} second class constraints are absent.
To obtain the total Hamiltonian density, according to the Dirac method \cite{Dirac}, we add to the density of the Hamiltonian the
Lagrange multiplier terms $v(x)\pi_0$, $u(x)\partial_m \pi^m$,
\begin{equation}
 {\cal H}_T=D_i E^i+{\cal F}\exp(-\beta{\cal F})+\pi^m \partial_m A_0+v(x)\pi_0+u(x)\partial_m \pi^m,
\label{16}
\end{equation}
where the functions $v(x)$, $u(x)$ do not possess physical meaning and are auxiliary variables that are connected with gauge degrees of freedom. The first class constraints in Eq. (16) generate the gauge transformations. Thus, Eq. (16) represents the set of Hamiltonians.
The physical space is the constant surface and one can get the energy density from the Hamiltonian on the constraint surface.
As a result, the density of energy, obtained from Eq. (16), is given by
\begin{equation}
\rho=D_i E^i+{\cal F}\exp(-\beta{\cal F}).
\label{17}
\end{equation}
The energy density (17) can be obtained also from the energy-momentum tensor $T_{\mu\nu}$ by the relation $\rho=T_0^{~0}$ \cite{Kruglov0}.
One must represent the total density of Hamiltonian (16) in terms of fields $A_\mu$ and momenta $\pi_\mu$ to obtain equations of motion.
Then we find the Hamiltonian equations
\begin{equation}
\partial_0 A_i=\{A_i,H\}=\frac{\delta H}{\delta \pi^i}= -E_i+\partial_i A_0-\partial_i u(x) ,
\label{18}
\end{equation}
\begin{equation}
\partial_0 \pi^i=\{\pi^i, H\}=-\frac{\delta H}{\delta A_i}=-\epsilon^{ijk}\partial_jH_k,
\label{19}
\end{equation}
\begin{equation}
\partial_0 A_0=\{A_0,H\}=\frac{\delta H}{\delta \pi^0}=v(x),\hspace{0.1in}
\partial_0 \pi_0=\{\pi_0,H\}=-\frac{\delta H}{\delta
A^0}=\partial_m \pi^m,
\label{20}
\end{equation}
where $H=\int d^3x {\cal H}_T$.
Equation (18) represents the gauge covariant form of equation for the electric field. Equation (19) is equivalent to the second equation in (2), and Gauss's law is the second constraint in this Hamiltonian formalism. As the function $u(x)$ is arbitrary, we may introduce new function $u'(x)=u(x)-A_0$ and the Hamiltonian (16), after the integration by parts to get the term $A_0\partial_m\pi^m$, will not contain the $A_0$. Thus, the component $A_0$ is not the physical degree of freedom. For a particular case $v(x)=\partial_0u'(x)$, one finds from Eqs. (18), (20) the relativistic form of gauge transformations $A'_\mu(x)=A_\mu(x) -\partial_\mu\Lambda(x)$, where $\Lambda(x)=\int dtu'(x)$. There are two arbitrary functions $u'(x)$, $v(x)$ in the general case. The Hamiltonian equations (18), (19) give the time evolution of physical fields that are equivalent to the Euler-Lagrange equations, and Eqs. (20) represent the time evolution of
non-physical fields $A_0$ and $\pi_0$ which are connected with the gauge degrees of freedom, and the variables $\pi_0$, $\partial_m \pi^m$ equal zero as constraints.

The dynamical variables $\hat{A}_i$ and $\hat{\pi}_i=-\hat{D}_i$ have in quantum theory the commutator
\begin{equation}
\left[\hat{A}_i(\textbf{x},t),\hat{D}_j(\textbf{y},t)\right]=-i\delta_{ij}\delta(\textbf{x}-\textbf{y}) \label{21}
\end{equation}
and the wave function $|\Psi\rangle$ obeys the Schr\"{o}dinger equation
\begin{equation}
i\frac{d|\Psi\rangle}{dt}=H|\Psi\rangle,
\label{22}
\end{equation}
and the equations as follows \cite{Dirac}:
\begin{equation}
\hat{D}_0|\Psi\rangle=0,~~~~\partial_m\hat{D}^m|\Psi\rangle=0,
\label{23}
\end{equation}
where $\hat{D}_0=- \hat{\pi}_0$.
As a result, the physical state is invariant under the gauge transformations.
Eqs. (23) give restrictions on the physical state $|\Psi\rangle$ which is gauge invariant. The physical fields $\textbf{E}$, $\textbf{B}$,
$\textbf{D}$, $\textbf{H}$  are represented by the Hermitian operators and do not depend on $A_0$ and they are invariants of the gauge  transformations.
One can apply for exponential electrodynamics the gauge fixing method which is beyond the Dirac's approach \cite{Hanson}, \cite{Henneaux}.

\section{NLED coupled with GR and magnetic black holes}

The action of our model of exponential electrodynamics in general relativity is
\begin{equation}
I=\int d^4x\sqrt{-g}\left[\frac{1}{2\kappa^2}R+ {\cal L}\right],
\label{24}
\end{equation}
where $R$ is the Ricci scalar, $\kappa^2=8\pi G\equiv M_{Pl}^{-2}$, $G$ is Newton's constant, and $M_{Pl}$ is the reduced Planck mass.
The Einstein and NLED equations follow from action (24)
\begin{equation}
R_{\mu\nu}-\frac{1}{2}g_{\mu\nu}R=-\kappa^2T_{\mu\nu},
\label{25}
\end{equation}
\begin{equation}
\partial_\mu\left[\sqrt{-g}F^{\mu\nu}\left(1-\beta {\cal F}\right)\exp(-\beta{\cal F}\right]=0,
\label{26}
\end{equation}
where the symmetric energy-momentum tensor of our model is given by \cite{Kruglov0}:
\begin{equation}
T^{\mu\nu}=\exp(-\beta{\cal F})\left[\left(\beta {\cal F}-1\right)F^{\mu\lambda}F^\nu_{~\lambda}+g^{\mu\nu}{\cal F}\right].
\label{27}
\end{equation}
It should be mentioned that Dirac-type magnetic monopole solution exists without sources in (26) \cite{Bronnikov}. Electrically charged black holes and corresponding solutions were considered in \cite{Kruglov11}.
Our goal here is to obtain the static magnetic black hole solutions to Eqs. (25) and (26).
It was shown in \cite{Bronnikov} that the invariant ${\cal F}$ compatible with the spherical symmetry, for pure magnetic field, is given by
 ${\cal F}=q^2/(2r^4)$, where $q$ is a magnetic charge. The spherically symmetric line element in this case is given by
\begin{equation}
ds^2=-A(r)dt^2+\frac{1}{A(r)}dr^2+r^2(d\vartheta^2+\sin^2\vartheta d\phi^2),
\label{28}
\end{equation}
where the metric function $A(r)$ reads
\begin{equation}
A(r)=1-\frac{2GM(r)}{r},
\label{29}
\end{equation}
and the mass function $M(r)$ is
\begin{equation}
M(r)=\int_0^r\rho(r)r^2dr=m-\int^\infty_r\rho(r)r^2dr.
\label{30}
\end{equation}
Here $m=\int_0^\infty\rho(r)r^2dr$ is the magnetic mass of the black hole, and the energy density for the case $\textbf{E}=0$, found from Eq. (17), is as follows:
\begin{equation}
\rho=\frac{q^2}{2r^4}\exp\left(- \frac{\beta q^2}{2r^4}\right).
\label{31}
\end{equation}
From Eqs. (30) and (31) we obtain the mass function
\begin{equation}
M(r)=\frac{q^2}{2}\int_0^r \frac{dr}{r^2}\exp\left(- \frac{\beta q^2}{2r^4}\right)=\frac{q^{3/2}}{2^{11/4}\beta^{1/4}}\Gamma\left(\frac{1}{4},\frac{\beta q^2}{2r^4}\right),
\label{32}
\end{equation}
where $\Gamma(s,x)$ is incomplete gamma function given by the expression
\begin{equation}
\Gamma(s,x)=\int_x^\infty t^{s-1}e^{-t}dt.
\label{33}
\end{equation}
From Eq. (32) we find the magnetic mass of the black hole
\begin{equation}
 m=M(\infty)=\frac{q^{3/2}\Gamma(1/4)}{2^{11/4}\beta^{1/4}}\simeq \frac{0.54q^{3/2}}{
\beta^{1/4}}.
\label{34}
\end{equation}
Taking into account Eqs. (29) and (32) one obtains the metric function
\begin{equation}
 A(r)=1-\frac{Gq^{3/2}}{2^{7/4}\beta^{1/4}r}\Gamma\left(\frac{1}{4},\frac{\beta q^2}{2r^4}\right).
\label{35}
\end{equation}
From Eq. (35) we find the asymptotic value of the metric function at $r\rightarrow \infty$
\begin{equation}
A(r)=1-\frac{2Gm}{r}+\frac{G q^{2}}{r^2}-\frac{\beta Gq^4}{10r^6}+{\cal O}(r^{-7}).
\label{36}
\end{equation}
The solution (36) is similar to the RN solution with some corrections in the order of ${\cal O}(r^{-6})$.
At the limit $r\rightarrow\infty$ the spacetime asymptotically becomes the Minkowski spacetime. If $\beta=0$  we arrive at
Maxwell's electrodynamics and solution (36) is the RN solution.
The asymptotic value of the metric function at $r\rightarrow 0$, obtained from Eq. (35), is given by
\begin{equation}
A(r)=1-\exp\left(-\frac{\beta q^2}{2r^4}\right)\left[\frac{Gr^2}{2\beta}-\frac{3G r^6}{4\beta^2q^2}+{\cal O}(r^{10})\right].
\label{37}
\end{equation}
Equation (37) shows that we have the regular black hole solution at $r\rightarrow 0$.
Introducing new variables $x=(2/(\beta q^2))^{1/4}r$ and $b=2^{3/2}\sqrt{\beta}/(qG)$, we can represent the metric function (35) as follows:
\begin{equation}
 A(x)=1-\frac{\Gamma\left(\frac{1}{4},\frac{1}{x^4}\right)}{bx}.
\label{38}
\end{equation}
One can find horizons by solving the equation $A(r)=0$. Internal Cauchy $x_-$ and event $x_+$ horizons for different parameters $b$ are given in Table 1.
\begin{table}[ht]
\caption{Internal Cauchy $x_-$ and event $x_+$ horizons}
\centering
\begin{tabular}{c c c c c c c c c c c}\\[1ex]
\hline \hline 
$b$ & 0.1 & 0.2 & 0.3 & 0.4 & 0.5 & 0.6 & 0.7 & 0.8 & 0.82 & 0.83\\[0.5ex]
\hline 
 $x_-$ & 0.8697 & 0.9618 & 1.0437 & 1.1274 & 1.2205 & 1.3334 & 1.4886 & 1.786 & 1.9314 & 2.1453\\[0.5ex]
\hline
 $x_+$ & 35 & 17 & 11 & 7.76 & 5.89 & 5.894 & 3.5 & 2.6593 & 2.406 & 2.1453\\[1ex]
\hline
\end{tabular}
\end{table}
The plot of the function $\Gamma\left(\frac{1}{4},\frac{1}{x^4}\right)/x$ is represented in Fig. 1.
\begin{figure}[h]
\includegraphics[height=3.0in,width=3.0in]{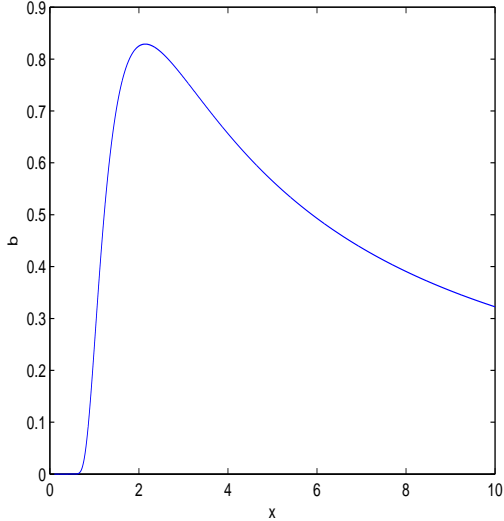}
\caption{\label{fig.1} The plot of the function $b=\Gamma\left(\frac{1}{4},\frac{1}{x^4}\right)/x$.}
\end{figure}
According to Fig. 1 there can be one, two or no horizons. At $b>0.83$ there are not horizons that lead to a particle-like solution.
 The function (38) for $b=1, 0.83, 0.5$ is given by Fig. 2.
\begin{figure}[h]
\includegraphics[height=3.0in,width=3.0in]{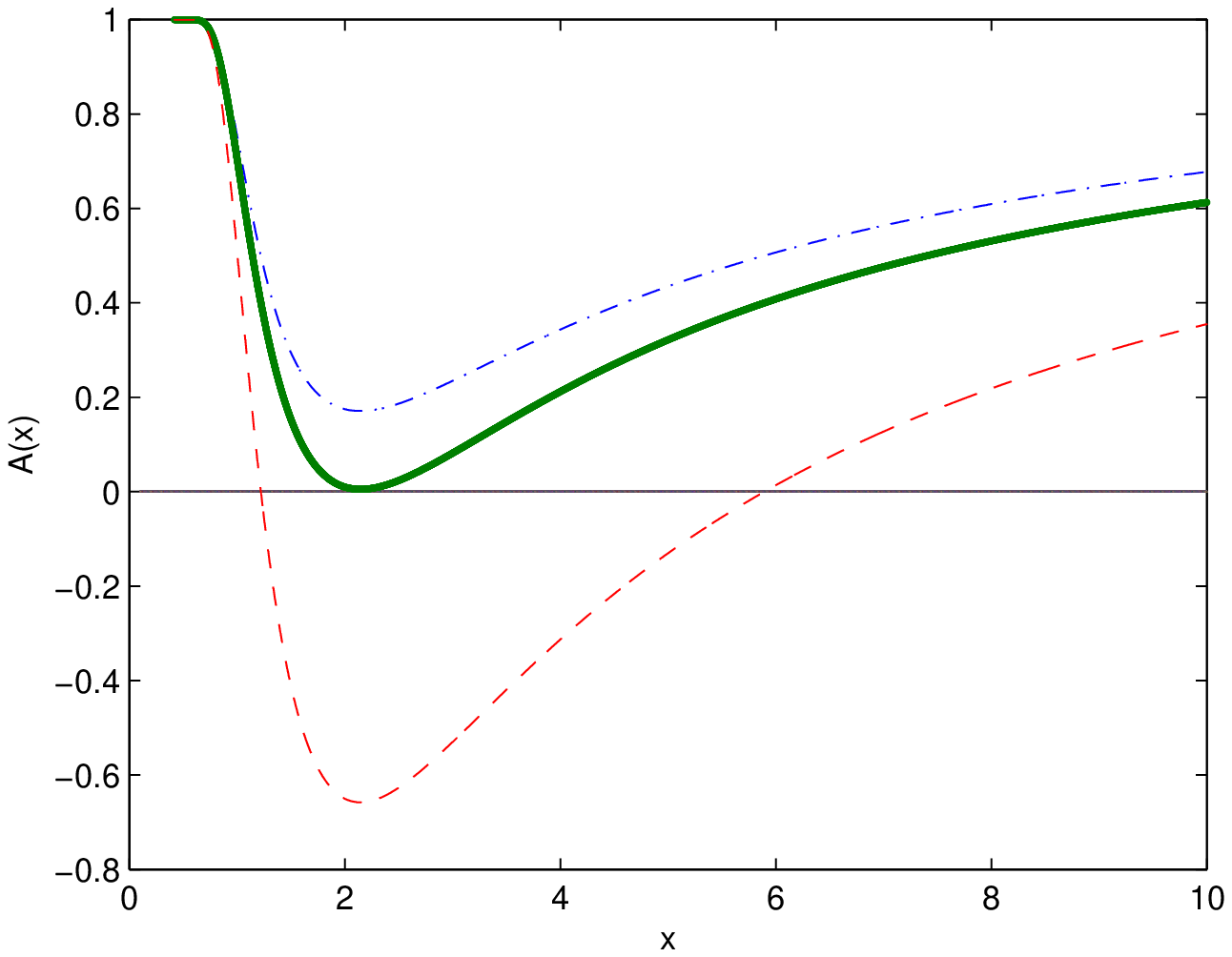}
\caption{\label{fig.2}The plot of the function $A(x)$. The dash-dot curve corresponds to $b=1$, the solid (thick) curve is for
$b=0.83$, and the dashed curve corresponds to $b=0.5$.}
\end{figure}
It follows from Fig. 2 that if $b\simeq 0.83$ there is one solution which corresponds to the extremal black hole. At $b<0.83$ we have two horizons of the regular black hole.

The trace of the energy-momentum tensor, obtained from (27), reads
\begin{equation}
{\cal T}\equiv T_{\mu}^{~\mu}=4\beta {\cal F}^2\exp(-\beta{\cal F})=\frac{\beta q^4}{r^8}\exp\left(-\frac{\beta q^2}{2r^4}\right).                                                               \label{39}
\end{equation}
We can obtain the Ricci scalar from Einstein's equation (25)
\begin{equation}
R=\kappa^2{\cal T}=\frac{\kappa^2\beta q^4}{r^8}\exp\left(-\frac{\beta q^2}{2r^4}\right).
\label{40}
\end{equation}
It follows from Eq. (40) that at $r\rightarrow \infty$ and at $r\rightarrow 0$ the Ricci scalar approaches to zero, $R\rightarrow 0$, i.e. there are no singularities of Ricci's curvature. However the Kretschmann scalar possesses the singularity only at $r=0$ (see Appendix). Therefore, at $r\rightarrow \infty$ spacetime
becomes flat. It should be mentioned that regular black hole solution in GR coupled to NLED was obtained in \cite{Beato}.
Let us consider the weak energy condition (WEC) \cite{Hawking} which guarantees that the energy density is positive for any local observer. For a system which obeys the spherical symmetry the radial magnetic field is $B(r)=F_{23}=-F_{32}$ and the components of the energy-momentum tensor are $\rho=T_0^{~0}=T_r^{~r}=-p_r$, where $p_r$ is a radial pressure. As a result $\rho+p_r=0$.
WEC reads: $\rho\geq 0$, $\rho+p_r\geq0$, $\rho+p_\perp\geq0$, where $p_\perp=-\rho-r\rho'/2$ and $\rho'=d\rho/dr$ \cite{Dymnikova}, \cite{Balart1}. The first two conditions are satisfied. Let us verify the third condition. From Eq. (31) we obtain
\begin{equation}
\rho'(r)=\frac{q^2(\beta q^2-2r^4)}{r^9}\exp\left(-\frac{\beta q^2}{2r^4}\right).
\label{41}
\end{equation}
Therefore, if $\rho'(r)\leq0$ then $p_\bot+\rho=-r\rho'/2\geq0$ and WEC will be satisfied.
We find from Eq. (41) that the requirement $\rho'(r)\leq0$ leads to $\beta{\cal F}=\beta q^2/(2r^4)\leq 1$. Thus, at $r\geq(\beta q^2/2)^{1/4}$ WEC is satisfied. The same restriction on the magnetic field, $\beta{\cal F}\leq 1$, was made from the causality principle.

\section{The black holes thermodynamics}

To study the thermal stability of charged black holes we will calculate the Hawking temperature and heat capacity of the black hole. If the Hawking temperature and heat capacity change the sign, this will indicate on the first-order phase transition. The point where the heat capacity is singular corresponds to the second-order phase transition \cite{Davies}. The unstable state of the black hole holds in the region of the negative temperature. The Hawking temperature is defined as follows:
\begin{equation}
T_H=\frac{\kappa_S}{2\pi}=\frac{A'(r_+)}{4\pi},
\label{42}
\end{equation}
where $\kappa_S$ is the surface gravity and $r_+$ is the event horizon. From Eqs. (28) and (29), one can find the relations
\begin{equation}
A'(r)=\frac{2 GM(r)}{r^2}-\frac{2GM'(r)}{r},~~~M'(r)=r^2\rho,~~~M(r_+)=\frac{r_+}{2G}.
\label{43}
\end{equation}
From Eqs. (30),(42), and (43) we obtain the Hawking temperature
\begin{equation}
T_H=\frac{1}{2^{7/4}\pi\sqrt{q}\beta^{1/4}}\left(\frac{1}{x_+}-\frac{4\exp(-1/x^4_+)}{x^2_+\Gamma\left(\frac{1}{4},\frac{1}{x^4_+}\right)}\right),
\label{44}
\end{equation}
where we took into account the relations
\begin{equation}
x_+=\left(\frac{2}{\beta q^2}\right)^{1/4}r_+ ,~~~bx_+= \Gamma\left(\frac{1}{4},\frac{1}{x^4_+}\right),~~~b\equiv \frac{2^{3/2}\sqrt{\beta}}{Gq}.
\label{45}
\end{equation}
The plot of the function $T_H\sqrt{q}\beta^{1/4}$ vs $x_+$ is represented in Fig. 3.
\begin{figure}[h]
\includegraphics[height=3.0in,width=3.0in]{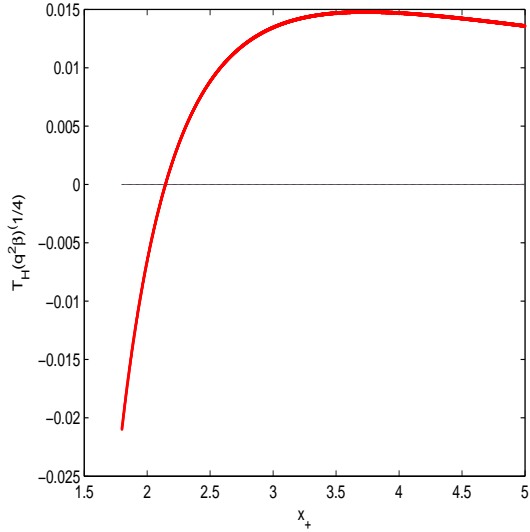}
\caption{\label{fig.3}The plot of the function $T_H\sqrt{q}\beta^{1/4}$ vs $x_+$.}
\end{figure}
At $x_+\simeq 2.145$ ($r_+\simeq 1.8\sqrt{q}\beta^{1/4}$) the temperature is zero, $T_H=0$ and, therefore, there is the black hole phase transition of the first-order. If $x_+< 2.145$ the Hawking temperature becomes negative that indicates that the black hole is unstable. The maximum of the temperature occurs at $x_+\simeq 3.733$ ($r_+\simeq 3.139\sqrt{q}\beta^{1/4}$). At this value $\partial T_H/\partial r_+=0$ and the heat capacity diverges which tells us that there is the phase transition of the second-order.
To study the heat capacity we use the entropy which satisfies the Hawking area law $S=Area/(4G)=\pi r_+^2/G$. Then we explore the heat capacity at the constant charge
\begin{equation}
C_q=T_H\frac{\partial S}{\partial T_H}|_q=\frac{T_H\partial S/\partial r_+}{\partial T_H/\partial r_+}=\frac{2\pi r_+T_H}{G\partial T_H/\partial r_+}.
\label{46}
\end{equation}
From Eqs. (44) and (46) we arrive at heat capacity
\begin{equation}
\frac{G}{\pi q\sqrt{2\beta}}C_q=\frac{x_+^7\Gamma\left(\frac{1}{4},\frac{1}{x_+^4}\right)^2e^{\frac{1}{x_+^4}}-4x_+^6\Gamma\left(\frac{1}{4},\frac{1}{x_+^4}\right)}
{(8x_+^4-16)\Gamma\left(\frac{1}{4},\frac{1}{x_+^4}\right)-x_+^5\Gamma\left(\frac{1}{4},\frac{1}{x_+^4}\right)^2e^{\frac{1}{x_+^4}}
+16x_+^3e^{-\frac{1}{x_+^4}}}.
\label{47}
\end{equation}
The plot of the function $C_q$ is represented by Fig. 4.
\begin{figure}[h]
\includegraphics[height=3.0in,width=3.0in]{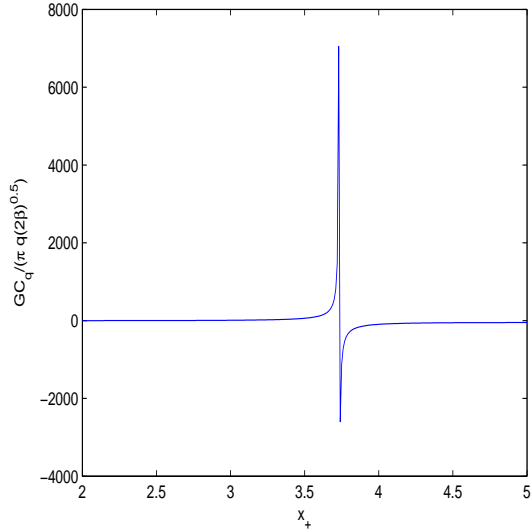}
\caption{\label{fig.4} The plot of the function $GC_q/(\pi q\sqrt{2\beta})$ vs $x_+$.}
\end{figure}
Heat capacity $C_q$ is singular because the denominator of $C_q$ becomes zero at the value $x_+\simeq 3.733$ as we mentioned before.
As a result, we have the second-order phase transition at $x_+\simeq 3.733$.
From Eq. (45) we find the constant $b\simeq 0.6845$ at the phase transition point. Fig. 4 shows that if $x_+\geq 3.733$
($r_+\geq 3.139 \sqrt{q}\beta^{1/4}$), the heat capacity is negative, $C_q<0$, and the black hole becomes unstable.
One can obtain the critical values of the
parameters $\beta$, $m$, and $T_H$ which correspond to the horizon, $x_+\simeq3.733$,
\begin{equation}
\beta=\frac{(bqG)^2}{8}\simeq 0.059 q^2G^2,~~~m=\frac{0.54 q^{3/2}}{\beta^{1/4}}\simeq \frac{1.1q}{\sqrt{G}},~~~
T_H\simeq \frac{0.094}{q\sqrt{G}}.
\label{48}
\end{equation}
For these values we obtain the critical horizon at $r_+ =\beta^{1/4}\sqrt{q}x_+/2^{1/4}\simeq 1.54q\sqrt{G}$.
Thus, the parameter of the model $\beta$, the magnetic mass of the black hole, and the Hawking temperature, corresponding to
the second-order phase transition, are expressed through the magnetic charge of the black hole $q$ and the Newton constant $G$.
As a result, if the horizon is greater than the critical value $r_+ \simeq 1.54q\sqrt{G}$ the black hole becomes unstable.
The first-order phase transition occurs at $x_+\simeq 2.145$ ($r_+\simeq 1.8\sqrt{q}\beta^{1/4}$). If $r_+< 1.8\sqrt{q}\beta^{1/4}$ the Hawking temperature is negative and the black hole is unstable.
We find the parameters corresponding to first-order phase transition
\[
 b=\frac{\Gamma (1/4,1/x_+^4)}{x_+}\simeq0.829,~~~~r_+\simeq 0.97q\sqrt{G},
\]
\begin{equation}
\beta\simeq 0.086 q^2G^2,~~~~m\simeq \frac{q}{\sqrt{G}}.
\label{49}
\end{equation}
For these parameters the Hawking temperature is zero, $T_H\simeq 0$.
As a result, the black hole, in the framework of our model, is stable for the range of the event horizons $0.97q\sqrt{G}<r_+<1.54 q\sqrt{G}$.

 \section{Conclusion}

We have considered exponential electrodynamics which is converted to Maxwell's electrodynamics for weak fields. In this model the birefringence
phenomenon takes place \cite{Kruglov0} similar to quantum electrodynamics with loop corrections \cite{Heisenberg}.
We have shown that WEC is satisfied if $ B\leq \sqrt{2}/\sqrt{\beta}$ that guarantees that the energy density is non-negative for any local observer.
The causality and unitarity principles are satisfied in our model at $ B\leq \sqrt{(5-\sqrt{17})/(2\beta)}$.
Because the parameter of the model $\beta\leq 1\times 10^{-23}$ T$^{-2}$
is very small \cite{Kruglov0} the causality principle and the unitarity of the theory take place up to very strong electromagnetic fields.
Thus, the model is of definite theoretical interest.
The gauge covariant Dirac's quantization of the exponential model of nonlinear electrodynamics was performed. This procedure of quantization is similar to classical electrodynamics and BI electrodynamics quantization.
We have investigated black holes possessing a magnetic charge in the framework of our model. The regular black hole
solution was obtained in general relativity and the asymptotic black hole solution at $r\rightarrow \infty$ was found.
We have calculated the magnetic mass of the black hole and the metric function. It was demonstrated that the Ricci scalar does not have singularities at $r\rightarrow \infty$ and at $r\rightarrow 0$. We shown that only at $b=2^{3/2}\sqrt{\beta}/(qG)\leq0.83$ there are two or
one horizons. If $b>0.83$ we have only a particle-like solution and no horizons.
The thermodynamic properties and thermal stability of regular black holes were analysed. We have calculated the
Hawking temperature of black holes and their heat capacity at the constant magnetic charge. It was shown that at the horizon $r_+\simeq 1.8\sqrt{q}\beta^{1/4}$ the first-order phase transition occurs. If $r_+<1.8\sqrt{q}\beta^{1/4}$ the black hole is unstable.
When heat capacity diverges, there is a phase transition of the second-order, which takes place at $r_+\simeq 3.139\sqrt{q}\beta^{1/4}$.
If the horizon $r_+> 3.139\sqrt{q}\beta^{1/4}\simeq 1.54q\sqrt{G}$ the black hole becomes unstable. We have estimated the parameters $\beta$, $m$ and $T_H$ corresponding to first-order and second-order phase transitions. We found the range, $0.97q\sqrt{G}<r_+<1.54 q\sqrt{G}$, when the  black hole is stable.

The results obtained in this paper show that the model of exponential electrodynamics
possesses attractive characteristics, may be tested in cosmology, and could describe physics of
black holes.

\section{Appendix}

Let us estimate the Kretschmann scalar $K(r)$ which can be calculated from the relation \cite{Hendi}
\begin{equation}\label{50}
 K(r)\equiv R_{\mu\nu\alpha\beta}R^{\mu\nu\alpha\beta}=A''^2(r)+\left(\frac{2A'(r)}{r}\right)^2+\left(\frac{2A(r)}{r^2}\right)^2,
\end{equation}
where $A'(r)=\partial A(r)/\partial r$, $x=(2/(\beta q^2))^{1/4}r$ and the metric function $A(x)$ is given by
\begin{equation}\label{51}
  A(x)=1-\frac{\Gamma\left(\frac{1}{4},\frac{1}{x^4}\right)}{bx}.
\end{equation}
Making use of the equality \cite{Stegun}
\begin{equation}\label{52}
  \frac{\partial \Gamma(s,z)}{\partial z}=-z^{s-1}\exp(-z),
\end{equation}
we obtain from Eqs. (51) and (52)
\begin{equation}\label{53}
  A'(x)=\frac{\Gamma\left(\frac{1}{4},\frac{1}{x^4}\right)}{bx^2}-\frac{4}{bx^3}\exp\left(-\frac{1}{x^4}\right),
\end{equation}
\begin{equation}\label{54}
  A''(x)=-\frac{2\Gamma\left(\frac{1}{4},\frac{1}{x^4}\right)}{bx^3}+\frac{16(1+x^4)}{bx^8}\exp\left(-\frac{1}{x^4}\right).
\end{equation}
With aid of Eqs. (53) and (54) and the equality $x=(2/(\beta q^2))^{1/4}r$ one can write down the Kretschmann scalar (50).
It is important to investigate the asymptotic of the Kretschmann scalar at $r\rightarrow 0$ and $r\rightarrow \infty$. For this purpose we explore the relations \cite{Stegun}
\begin{equation}\label{55}
\Gamma(s,z)=\Gamma(s)-z^s\left[\frac{1}{s}-\frac{z}{s+1}+\frac{z^2}{2(s+2)}+{\cal O}(z^3)\right]~~~~z\rightarrow 0,
\end{equation}
\begin{equation}\label{56}
\Gamma(s,z)=\exp(-z)z^s\left[\frac{1}{z}+\frac{s-1}{z^2}+\frac{s^2-3s+2}{z^3}+{\cal O}(z^{-4})\right]~~~~z\rightarrow\infty.
\end{equation}
From Eqs. (55) and (56) we obtain the asymptotic
\begin{equation}\label{57}
\Gamma \left(\frac{1}{4},\frac{1}{x^4}\right)=\exp\left(-\frac{1}{x^4}\right)\left[x^3-\frac{3}{4}x^7+{\cal O}(x^{11})\right]~~~~x\rightarrow 0,
\end{equation}
\begin{equation}\label{58}
\Gamma \left(\frac{1}{4},\frac{1}{x^4}\right)=\Gamma\left(\frac{1}{4}\right)-\frac{1}{x}\left[4-\frac{4}{5x^4}+{\cal O}(x^{-8})\right]~~~~z\rightarrow\infty.
\end{equation}
By virtue of Eqs. (51), (53), (54) and (58) one finds
\begin{equation}\label{59}
  \lim_{x\rightarrow\infty}\frac{A(x)}{x^2}=\lim_{x\rightarrow\infty}A'(x)=\lim_{x\rightarrow\infty}A''(x)=0.
\end{equation}
Then from Eq. (50) and the relation $x=(2/(\beta q^2))^{1/4}r$ we obtain
\begin{equation}\label{60}
   \lim_{r\rightarrow\infty}K(r)=0.
\end{equation}
Thus, Eq. (60) shows that there is no the singularity of the Kretschmann scalar at $r\rightarrow \infty$.
Making use of Eqs. (51), (53), (54) and (57) one can find
\begin{equation}\label{61}
  \lim_{x\rightarrow 0}\frac{A(x)}{x^2}=\infty,~~~ \lim_{x\rightarrow 0}\frac{A'(x)}{x}=\lim_{x\rightarrow 0}A''(x)=0.
\end{equation}
As a result,
\begin{equation}\label{60}
   \lim_{r\rightarrow 0}K(r)=\infty,
\end{equation}
and the Kretschmann scalar has the singularity at $r=0$. This also occurs in other models of NLED \cite{Hendi}.

\end{document}